\documentclass[conference]{IEEEtran}
\IEEEoverridecommandlockouts

\usepackage{cite}
\usepackage{amsmath,amssymb,amsfonts}
\usepackage{algorithmic}
\usepackage{graphicx}
\usepackage{textcomp}
\usepackage{xcolor}
\usepackage{hyperref}

\def\BibTeX{{\rm B\kern-.05em{\sc i\kern-.025em b}\kern-.08em
    T\kern-.1667em\lower.7ex\hbox{E}\kern-.125emX}}
\begin{document}

\title{Trustworthy image-to-image translation: evaluating uncertainty calibration in unpaired training scenarios\\
}
\author{\IEEEauthorblockN{1\textsuperscript{st} Ciaran Bench}
\IEEEauthorblockA{\textit{Department of Data Science and AI} \\
\textit{National Physical Laboratory}\\
Teddington, United Kingdom \\
ciaran.bench@npl.co.uk}
\and
\IEEEauthorblockN{2\textsuperscript{nd} Emir Ahmed}
\IEEEauthorblockA{\textit{Department of Data Science and AI} \\
\textit{National Physical Laboratory}\\
Teddington, United Kingdom \\
emir.ahmed@npl.co.uk}
\and
\IEEEauthorblockN{3\textsuperscript{rd} Spencer Thomas}
\IEEEauthorblockA{\textit{Department of Data Science and AI} \\
\textit{National Physical Laboratory}\\
Teddington, United Kingdom \\
spencer.thomas@npl.co.uk}
}
\maketitle

\begin{abstract}
Mammographic screening is an effective method for detecting breast cancer, facilitating early diagnosis. However, the current need to manually inspect images places a heavy burden on healthcare systems, spurring a desire for automated diagnostic protocols. Techniques based on deep neural networks have been shown effective in some studies, but their tendency to overfit leaves considerable risk for poor generalisation and misdiagnosis, preventing their widespread adoption in clinical settings. Data augmentation schemes based on unpaired neural style transfer models have been proposed that improve generalisability by diversifying the representations of training image features in the absence of paired training data (images of the same tissue in either image style). But these models are similarly prone to various pathologies, and evaluating their performance is challenging without ground truths/large datasets (as is often the case in medical imaging). Here, we consider two frameworks/architectures: a GAN-based cycleGAN, and the more recently developed diffusion-based SynDiff. We evaluate their performance when trained on image patches parsed from three open access mammography datasets and one non-medical image dataset. We consider the use of uncertainty quantification to assess model trustworthiness, and propose a scheme to evaluate calibration quality in unpaired training scenarios. This ultimately helps facilitate the trustworthy use of image-to-image translation models in domains where ground truths are not typically available.
\end{abstract}


\section{Introduction}

Over 2.3 million new cases of breast cancer were diagnosed in 2020, making it one of the most prevalent cancers worldwide \cite{arnold2022current}. Successful treatment often hinges on the early detection of the disease, where the application of radiation, chemotherapies, or surgery are more likely to result in positive patient outcomes \cite{carlson2003treatment}. However, screening is usually carried out via manual inspection of images, placing considerable strain on healthcare systems and restricting its use to regions where there are suitably trained staff. Therefore, there remains considerable desire for computational methods that can automate the screening process, expanding access to this life-saving service.

Algorithms based on deep neural networks have been shown particularly effective at learning the screening task, matching or exceeding the performance of human radiologists \cite{mckinney2020international}. A great deal of their success may be attributed to their capabilities for automatic feature extraction; models can learn to i) detect the most useful features and ii) combine and/or transform them to produce accurate predictions \cite{paul2016deep}. In contrast, non-network based methods tend to involve the use of manually extracted features derived using assumptions or prior knowledge about the prediction task that may be incomplete or inaccurate. These are then fed into models that may not have sufficient descriptive capacity to perform adequate nonlinear transformations. Furthermore, given the variation in the properties of the images acquired across different populations/locations, it is non-trivial to devise consistent methods to extract these hand-engineered features. In contrast, deep networks can cope with some of this variation implicitly provided they are given a diverse enough training dataset.

\subsection{Model generalisability}
Despite several reports describing impressive predictive capabilities, deep networks are not widely used for disease detection in the clinic \cite{mckinney2020international}. To facilitate widespread use, there is a need to establish the trustworthiness of a model's performance on unseen images given their tendency to overfit to training data. But generic implementations of deep networks are incapable of providing robust information about trustworthiness (while generic classifiers provide confidence scores, these are typically overestimated and do not reflect the true underlying doubt in a given prediction \cite{bai2021don}). Furthermore, it is not straightforward to interpret the behaviour of deep networks, making it challenging to determine whether the model is making its prediction using clinically relevant information. These factors make models trained with generic supervised optimisation schemes unsuitable for clinical use.

Several strategies have been proposed to make the predictions by deep networks more trustworthy. Explainable AI is one theme of research where trustworthiness is improved by providing a means to interpret how a model derives its predictions \cite{saranya2023systematic}. For example, this includes techniques based on the use of saliency maps that reveal the information a model may be using when it makes a prediction from a given input \cite{pertuz2023saliency}. One can use this to establish trustworthiness by assessing whether the model is using features from the most clinically relevant regions of an input image as opposed to other regions that might indicate the model is overfitting (such as information from the non-tissue background). 

Other work has focused on improving a model's trustworthiness by enhancing and assessing its generalisability \cite{mckinney2020international,oza2022image}. Poor generalisation occurs due to the presence of a domain gap between a model's training data and test data. Note that in `test data' we also include data from sources separate to those used to evaluate test performance in a typical training/testing framework, e.g. data from a different hospital once the model is `deployed'. Differences in the frequency of occurrence of various features, the relationship between these features and the ground truth, as well as the types of features themselves may contribute to this domain gap. 

More precisely, a data domain $\mathcal{D} = \{\chi, P(x), P(x,y)\}$ consists of an input feature space $\chi$ (a vector space containing all image features), a marginal distribution $P(x)$, and a joint probability distribution $P(x,y)$, where $x$ is an instance of the set $\textbf{x}$ of $N$ total network inputs $x_1, x_2, ... x_N \in \textbf{x}$ and $y$ is an instance of the corresponding set of ground truths $\textbf{y}$ $y_1, y_2, ...y_N \in \textbf{y}$ \cite{kouw2019review}. Poor generalisability occurs when there is a large discrepancy between $\mathcal{D}_{\text{train}}$ and $\mathcal{D}_{\text{test}}$.

One approach to reduce the domain gap is to expand the training dataset so that the model is less likely to encounter new, unseen features (i.e. expand the training data distribution so that it is in greater alignment with that of any test data). Ideally, this is done by adding more relevant examples to the training set. However in some cases this additional data may not be available. Instead, additional examples may be produced by modifying existing training data with various transformations in a process known as data augmentation \cite{oza2022image}. Generic transformations originally devised for models trained on natural images have been applied to medical images, e.g. rotation, rescaling, flipping, and shifting \cite{costa2019data}. However, such generic transformation may not capture the new representations of tissue features that emerge as a consequence of using a different X-ray scanner, or differences in feature distributions across various patient populations \cite{oza2022image}. Moreover, the generic transformations must be semantically modified for a type of input (e.g. images) and are restricted in terms of transformations and parameterisation \cite{ntelemis_generic_2024}. Image-to-image translation (a form of neural style transfer) can been used as a means to expand the training data distribution to contain these less generic feature representations.

\subsection{Style transfer as data augmentation}
Style transfer tasks involving images (i.e. image-to-image translation) usually involve images from different domains, a source domain $\mathcal{D}_{\text{source}}$, and a target domain $\mathcal{D}_{\text{target}}$. A network $G: \mathcal{D}_{\text{source}} \rightarrow \mathcal{D}_{\text{target}}$ takes an image $x_s\in \mathcal{D}_{\text{source}}$ as an input and transforms it such that the output $\hat{x}_s$ appears to be described by $\mathcal{D}_{\text{target}}$, while still retaining much of the image's original structure \cite{hoyez2022unsupervised}. 

In the case of mammography images $\hat{x}_s$ should depict the same tissue as shown in $x_s$, but represented in a manner such that it appears to have been acquired from the scanner/processing settings used to acquire images belonging to $\mathcal{D}_{\text{target}}$. However, this is challenging given that some populations may have structures not commonly found in another population, e.g. higher instances of very low tissue densities in western populations compared to East Asian populations \cite{mariapun2015ethnic}. In this case, there are few examples exhibiting the kinds of features that typically represent these structures in the target style. A model may be prone to hallucinate features in this scenario. Therefore, the ideal behaviour is that any structures/objects shared in each datsaet should have the same feature representation, while those that are not should be preserved or only mildly adapted. 

There are other complications emerging from the use of the technique, e.g. that the adaptation only helps align the feature spaces without considering the joint distributions of each domain, which may result in the introduction of unrealistic relationships between the presence of some features and a given disease class. Nevertheless, this approach has been used effectively in various mammography studies, where style transfer may take the form of inserting lesions into images of otherwise healthy tissues \cite{wu2018conditional}, or imposing the style of different scanners onto whole images \cite{wang2020mr}, and  improving the generalisability of models trained on downstream tasks.

\subsection{Unpaired image-to-image translation models}
Several supervised approaches to style transfer have been proposed for medical image applications \cite{mcnaughton2023machine, jung2018inferring, fernandez2021improving}. However, large scale mammography datasets composed of paired images (i.e. images of the same patient acquired with different scanners) are not typically available. Acquiring them would take significant effort from a resource limited workforce, involving  uncomfortable procedures and radiation exposure for the patients involved. This motivates the use of unpaired style transfer techniques.

Here, we consider two commonly used frameworks: GANs, specifically, cycle consistent generative adversarial networks (cycleGANs) \cite{zhu2017unpaired} and a diffusion-based SynDiff model \cite{ozbey2023unsupervised}. We treat the cycleGAN as our baseline given its widespread use in medical imaging tasks. We also consider the diffusion-based SynDiff model which, unlike cycleGANs, directly optimises a correlate of the data likelihood and uses multi-step generation/adaptation. 

\subsubsection{cycleGAN}
CycleGANs are a commonly used architecture for unpaired image-to-image translation \cite{zhu2017unpaired}. They are composed of four module networks; two generators ($G_A$ and $G_B$) and two discriminators ($D_A$ and $D_B$) that process two sets of images; one described by a data domain $\mathcal{D}_A$, and another described by $\mathcal{D}_B$ where there is some overlap in terms of the semantic content of the images, but generally $\mathcal{D}_A \neq \mathcal{D}_B$. Here, $G_A$ will be trained to take an image from $\mathcal{D}_B$ and transform it such that it appears more like it belongs to $\mathcal{D}_A$. $G_B$ transforms images from $\mathcal{D}_A$ to appear more like they belong to $\mathcal{D}_B$. The discriminators assess whether an image is originally from a particular domain (e.g. $\mathcal{D}_A$ for $D_A$), or whether it has been transformed to appear like images from that domain. 

\subsubsection{SynDiff}
Diffusion models are a different framework for learning image generation, where initially a network is trained to iteratively add noise to an image until its contents are indistinguishable from random noise (modelled as a diffusion process using Markov chains). Image generation is performed by learning to reverse this process, i.e. iteratively remove noise from an grid of random noise to form an image \cite{yang2023diffusion}.

Unlike GANs, which do not optimise over an explicit evaluation of the data likelihood (i.e. how well the model explains the observed training data) as part of the training process, diffusion models optimise over a correlate of the data likelihood. This can be a more effective loss compared to the indirect/implicit optimisation of the likelihood associated with generic GANs. Furthermore, with a diffusion model the transformation of images is executed over multiple steps rather than a one-shot approach, providing a means to gradually refine the images which can improve the quality of outputs.

The generic diffusion model framework was modified in \cite{ozbey2023unsupervised} to learn efficient, unpaired, cycle-consistent, image-to-image translation (SynDiff). The basic idea is that after training, the model should be able to take a given source image as an input along with an image of pure noise, and denoise the latter in only a few steps to generate our style transferred image (i.e. the source image styled so it appears to belong to the target domain). This denoising model is akin to a conditional generator, referred to as a source conditional adversarial projector, which is trained with a corresponding discriminator. 
\subsection{Evaluating performance with uncertainty quantification}
The most direct way to assess whether the augmentation performed by a style transfer model is effective is to observe whether the performance/generalisability of the model improves with the inclusion of augmented data in its training set \cite{cha2020evaluation}. But this may be prohibitively expensive to implement, as several iterations of a style transfer model may need to be trained to achieve suitable performance. This motivates our focus on using comparatively more efficient uncertainty quantification and other quality metrics that can be applied to the style transfer models themselves (or their outputs), avoiding the need to train a subsequent model to evaluate performance. Uncertainty quantification provides a means to assess the degree of doubt in a given prediction. It is often argued that it should correlate with prediction error, allowing one to determine when a prediction is likely to be correct \cite{pernot2023calibration}. We consider its application on both the cycleGAN and SynDiff models to cover the two most commonly implemented frameworks for neural style transfer.

We opt for Monte Carlo Dropout on the cycleGAN given the strong precedent for its use on the architecture and due to its computational efficiency \cite{kendall2017uncertainties}. There is currently no known Bayesian variant of SynDiff. Therefore we apply deep ensembles, which is model-agnostic (i.e. doesn't require changes to the optimisation scheme), and is similarly straightforward to implement, but more expensive.
\subsubsection{Monte Carlo Dropout}
Several works have estimated the predictive variance of a cycleGAN by evaluating a given input several times with dropout layers in the generator left active \cite{upadhyay2021uncertainty,upadhyay2021robustness,karthik2023uncertainty}. It has been shown that for non-generative supervised models this technique approximates variational inference, allowing one to estimate predictive uncertainty \cite{kendall2017uncertainties}. However, whether the use of dropout during the training and evaluation of a cycleGAN approximates the use of variational inference has not been assessed, putting the validity of any uncertainty estimates produced with the technique into question. With that said, a scheme for Monte Carlo dropout has been derived for generic GANs, where Palakkadavath et al. \cite{palakkadavath2021bayesian} argue that the application of dropout and weight regularisation to the generator and discriminator, along with the addition of a weight regularisation term to the loss is sufficient to implement the technique (assuming the prior on the parameters of the generator is a Gaussian distribution). The authors of \cite{tiao2018cycle} and others have derived variational inference schemes for optimising cycleGAN models, but their potential equivalence with a corresponding dropout objective have not been rigorously assessed. A Bayesian formulation of a cycleGAN was proposed  in \cite{you2020bayesian,RanWang2022}, where it is also mentioned that the use of a dropout objective (specifically applying dropout to the generators) can be used to construct a comparable framework. But this is not used in the context of uncertainty quantification, nor is this rigorously proved. Despite the lack of rigorous theoretical justification, the use of Monte Carlo dropout-inspired training/evaluation schemes on cycleGANs have been shown to produce estimates of predictive variance that correlate well with accuracy \cite{galapon2024feasibility}. 

There are several sources of uncertainty, but the two most dominant in practical imaging scenarios are aleatoric uncertainty (irreducible uncertainty intrinsic to the data) and epistemic uncertainty (reducible model uncertainty). While this implementation of Monte Carlo Dropout is commonly used in the context of cycleGANs, it does not explicitly model aleatoric uncertainty. The resultant uncertainty estimates are generally interpreted as epistemic uncertainty \cite{kendall2017uncertainties}. However, aleatoric uncertainty is likely to be a significant factor affecting model performance. Formulating an approach to modelling aleatoric uncertainty with the cycleGAN objective is left for future work, and in any case our aim is to assess the efficacy of the most widely used implementation of the technique in the context of style transfer. Therefore, we use this more generic implementation in our work.

\subsubsection{Deep ensembles}
We also investigate the use of uncertainty quantification for SynDiff. There are several reported approaches to performing uncertainty quantification on diffusion models \cite{ kou2023bayesdiff,xu2024bayesian,neumeier2024reliable}. However, most of these approaches were formulated for specific architectures/training tasks and it is unclear how they might be implemented for the SynDiff architecture and/or the case of conditional image generation. With that said, not all uncertainty quantification techniques require modifications to a given model's training scheme. Deep ensembles is one approach \cite{lakshminarayanan2017simple}, where several versions of the same model/architecture are trained on identical datasets, each time using a different weight initialisation. Each test example is then fed through each model in the ensemble, where the predictive variance may be interpreted as an uncertainty.

This approach has been used on diffusion models in various forms, e.g. in \cite{shu2024zero}, an ensemble of conditional diffusion models is trained for estimating predictive uncertainty for a regression task. In \cite{berryshedding} the computational expense of training multiple diffusion models is reduced by instead ensembling over a sub-module of the model. Ekmekci et al. \cite{ekmekci2023quantifying} also trained a generic ensemble of diffusion models, and subsequently decomposed the total uncertainty into aleatoric and epistemic components. Chan et al. \cite{chan2024hyper} implemented an efficient form of ensembling using a hyper-network framework as a conditional diffusion model, and similarly decompose the various sources of uncertainty.

Given this work is primarily focused on more practically straightforward methods, we implement the generic version of the ensembling approach by training five diffusion models on our chosen task.

\subsection{Evaluating uncertainty calibration}
Uncertainty estimates are considered accurate or `calibrated' if their magnitude correctly encodes the degree of doubt in a given prediction. Metrics for assessing the calibration of estimated uncertainties often compare whether the magnitude of the uncertainties correlate with prediction error/model accuracy. However, assessing model accuracy in the context of unpaired style transfer is non-trivial. As will be discussed, the Fr\'echet Inception Distance (FID) is one metric widely used as an accuracy metric \cite{heusel2017gans}. Though, it is widely known that it is less effective for non-natural images, such as mammography scans \cite{deshpande2024report}.

Additionally, these calibration metrics are often conditioned on the  magnitude of the estimated uncertainties, where the mean uncertainty and accuracy of uncertainty bins are compared \cite{pernot2023calibration}. However, the FID is only effective when large numbers of images are considered. Practically, this means the test set should be composed of tens of thousands of images, to form bins with sizable sample populations. But in many applications (including mammography) data is scarce, making this approach to assessing calibration using the FID as a proxy for accuracy impractical.

We propose a scheme to tackle these challenges. We demonstrate whether the FID is a suitable metric for accuracy by augmenting test data in a way that should degrade performance, and then observing whether the FID correlates with strength of the augmentation. If a strong correlation is observed, we may indirectly assess the calibration of estimated uncertainties by observing whether the strength of the augmentation corresponds to changes in the uncertainties produced from each evaluation. The use of augmented versions of the same test set provides a data-efficient means to assess conditional calibration.

\section{Methods}
\subsection{Data}
\label{sec:data}
We consider three open source datasets: VinDr Mammo (VDM) \cite{nguyen2023vindr}, the Chinese Mammography Database (CMMD) \cite{cai2023online}, and the Curated Breast Imaging Subset of the Digital Database for Screening Mammography (CBIS-DDSM) \cite{sawyer2016curated}. These were chosen based on 
the number of images/cases available, the relative homogeneity in image size, the varied patient populations (VDM and CMMD are composed of patients from Asian populations, while CBIS-DDSM is composed of patients from the US), scanners, and image mediums (CBIS-DDSM images are scanned film, while CMMD and VDM are natively digital). The variation in features due to these differences in population, image media, and scanner types impose a strong challenge on the chosen style transfer algorithms.

In order to alleviate the memory challenges of computing on whole mammography images we use a patch based approach for our experiments. Image patches were acquired by taking a preprocessed whole mammography image, and parsing 256$\times$256 pixel sections in steps of 246 pixels (producing an overlap of 10 pixels). This study only concerns filled patches, so only those where more than 99 \% of pixels were non-zero were considered for training/evaluation.

For VDM and CBIS-DDSM, the breast was segmented from the background using Otsu's method \cite{ostu1979threshold}, and contrast inversion was applied to images with photomometic interpretation set to \verb |MONOCHROME1|. All `right' laterality images were flipped horizontally. The amplitude of each image was normalised to a minimum of 0, and a maxium of 1. The whole images were padded to ensure they had dimensions of 2224$\times$2224 pixels to ensure they were compatible with the patch-parsing function. If a patch met our inclusion criteria, histogram equalisation was applied, followed by a subsequent normalisation. All CMMD images were pre-segmented and did not require this procedure. All other preprocessing steps were applied to the CMMD images. We set aside 1500 patches for training, and 3500 for testing from each set of images. Training patches were acquired by parsing all relevant patches from one set of whole images, while the test patches were acquired by parsing all patches from a separate set of whole images. Therefore, there is no overlap across the training/test sets at the patch or whole image level. The number of training patches was chosen so that SynDiff could be trained for 100 epochs within 4 days  using the available computational resources (NVIDIA A100 GPU, and a AMD EPYC 7643 2.3 GHz CPU).

To demonstrate that our methods generalise to non-medical images, we also consider the sketch-to-shoe image translation task described in \cite{zhu2017unpaired}. We use the original sizes of the images and clean the data set to remove paired examples and to prevent instances of `approximate' paired examples from occurring in the data. For the latter we remove colour variants of the same shoe (that would have approximately the same sketch) and horizontally flip half of the remaining examples. The only additional preprocessing performed is the addition of noise to test images for some experiments assessing the calibration of predicted uncertainties. We train on 1500 examples, and test on 3500 examples from each domain.

\subsection{Training/architectural configuration}
It is non-trivial to devise a sensible stopping criterion for generative models without a ground truth \cite{saad2024early}. Therefore, we trained the models in accordance with the number of epochs used in similar studies (e.g. 100 for SynDiff \cite{ozbey2023unsupervised}, and 200 for the cycleGAN \cite{zhu2017unpaired}), retaining the default architecture associated with each technique given the size of our images (256$\times$256 pixels) matched to those used in the default implementation of each model. Given the length of time required to train each SynDiff  model, an extensive hyperparameter search was not possible. With that said, the cycle-consistency loss was found to have the greatest effect on image quality as measured by the FID on 3500 test images. Here, a cycle consistency weighting of 1000 was used to train all SynDiff models. In contrast, the default hyperparameters of the cycleGAN were found to provide comparable performance (as measured by the FID on 3500 test images), and so no further experiments were conducted. In any case, while we do compare the metrics produced from the outputs of each model, a rigorous comparison of either architecture is beyond the scope of this work. Rather, we provide these results to demonstrate that both provide comparable performance for the follow-up studies in uncertainty quantification for one task, and to note that both models are capable of providing improvements in FID scores without extensive hyperparameter searches for a range of tasks.

\subsection{Evaluation metrics}

Our aim is to impose the style of the target domain onto source domain images, while preserving as much structural content as possible (i.e. ensuring the original tissue is still being represented in the adapted image). Therefore, we decompose the evaluation into two parts: assessing how well the target style has been imposed onto the source domain images, and assessing how much of the structural content has been preserved post-adaption. We assess adaption quality via metrics commonly used in the context of image generation, such as the FID for assessing the imposition of the target style, and the Continuous Wavelet Structural Similarity Index for content preservation \cite{cwssim,ozbey2023unsupervised,graf2023denoising}. It is well known that the FID is sub-optimal for medical images, in particular, because the Inception-Net filters are not especially sensitive to the features that define images \cite{deshpande2024report}. However, we retain its use here in the absence of any medical imaging specific metric and given the use of FID in comparable studies \cite{ozbey2023unsupervised,liang2022image,abdusalomov2023evaluating}.

The Structural Similarity Index (SSIM) is widely used in medical imaging studies \cite{7404021, super-resolution, Gan-baseddataaugmentation}
as it considers contextual information often important to defining structural features, and makes it better equipped to detect distortions which typically span regions of pixels \cite{1284395}. With that said, it is known to be highly sensitive to geometric and scale distortions \cite{7404021}. We observe that SynDiff may apply small offsets (e.g. 1-3 pixels) to images. So instead, we employ the Continuous Wavelet variant of SSIM (CWSSIM), which is less sensitive to such offsets and other distortions \cite{cwssim}.
When calculating the CWSSIM, all pixel values of the images were normalised to have the range [0,1], and then converted to 8-bit unsigned integers with a range of 0-255. 

\subsection{Uncertainty quantification and calibration}
For the mammography experiments, we consider six style transfer tasks across the three datasets. A baseline FID score for each task was computed using the unadapted images from each source and target domain. The FID was then calculated for each task using the style transferred version of these source images, along with the same set of target images. The CWSSIM was evaluated for each pre and post-adapted test image pair, along with the mean score over the whole test set for each task. Images were normalised to have a range from 0-1 before computing these metrics.

\subsubsection{Monte Carlo Dropout for cycleGAN}
\label{sec:mcd}
We trained a cycleGAN with dropout (20 \%) applied to several convolutional layers in the generator modules (using the default implementation \cite{zhu2017unpaired}). For evaluation, dropout is left active, and each input image is evaluated 25 times. While more samples may ultimately provide better performance (implicitly resulting in a more thorough sampling of the model's posterior distribution), here we use 25 Monte Carlo samples given data storage limitations. In other words, each input image $x_i$ is used to produce 25 corresponding style transferred outputs $\hat{x}_{i,m}$, where $m = 1,\dots,25$. The three RGB channels are averaged into a single channel, and the pixel-wise standard deviation is computed for each resultant image, $\sigma_i$, where the subsequent mean of all of its values $\mu^{\sigma}_{i}$, which we refer to as the predictive standard deviation (PSD) represents the uncertainty in the model's prediction \cite{galapon2024feasibility}. $\bar{\mu} = \frac{1}{3500}\sum^{3500}_{i=1}\mu^{\sigma}_{i}$ represents the mean uncertainty over the whole test set (mPSD). The raw model outputs are used in the calculation of $\bar{\mu}$, without any additional normalisation.

The ideal behavior of a style transfer algorithm is to ensure that all common structural objects are represented by the same feature representations. Higher order statistical measures comparing activation distributions (e.g. the FID) are used to assess the degree of alignment between domains. 
If one can make the assumption that this metric is a robust indicator of model correctness, then whatever metric is used as an indicator of uncertainty should ideally correlate with its values.

To this end our evaluation procedure involves evaluating a model on several versions of a test set, where each version is created by gradually adding larger amounts of noise. We assume that our model's performance will not be robust to images augmented with noise, and that this will have some kind of effect on both accuracy and the estimated uncertainties (as is reported here \cite{buzuti2023frechet} for accuracy). The suitability of the FID as an accuracy metric is coarsely assessed by observing whether the FID acquired from evaluating each test set once increases as increasing amounts of noise are added to the test images. 

We then aim to assess whether changes in this FID and the mPSD correlate under these circumstances. If the first experiment reveals that the FID does indeed correlate with accuracy, this subsequent experiment may provide an indication of whether uncertainty correlates with accuracy. 

While the precision of our assessment of this correlation is low given we have not precisely validated how well the FID correlates with the true underlying accuracy, this nonetheless provides some indication about whether we may interpret the PSD as an uncertainty. We set the range of each image to span 0 to 255, and then add noise with mean 0, and a variance of some percentage of the max value. We use this protocol for both the mammography and natural image experiments.

We also acknowledge that for all experiments considered, the choice of dropout rate can have a considerable effect on the chosen uncertainties, and that the observations seen here could differ significantly for a model trained and evaluated with different hyperparameters.

\subsubsection{Deep ensembles for SynDiff}
We train five identical models on the same training set as described in Section \ref{sec:data}, each with a random weight initialisation. We evaluate each test example, and compute the pixel-wise standard deviation to derive an expression of uncertainty. The same procedure described in Section \ref{sec:mcd} is used here for assessing calibration quality, where each augmented test image is fed to the five trained models to produce a predictive distribution. While the use of more models may improve results (providing a more thorough sampling of the model's posterior distribution \cite{fort2019deep}), here we train five due to large training times for the SynDiff model. SynDiff was found to apply a subtle (few pixels) offset to translated images. Therefore, we registered the outputs for each model to their respective source images using the ANTsPy library registration function in translation mode (\url{https://github.com/ANTsX/ANTsPy}). We cropped 5 pixels from each side of the translated images to ensure all images had the same dimensionality before computing the PSD. We plot the mean  FID computed from the outputs for each model, against the mPSD. 

\section{Results}
\subsection{cycleGAN: Non-medical images}

\begin{figure}
    \centering
     \hspace{1cm}{\fontfamily{cmss}\selectfont \large CycleGAN: Shoes $\leftrightarrow$ Sketches }
    \includegraphics[trim={0 0 0 1.8cm},clip,width=.9\columnwidth]{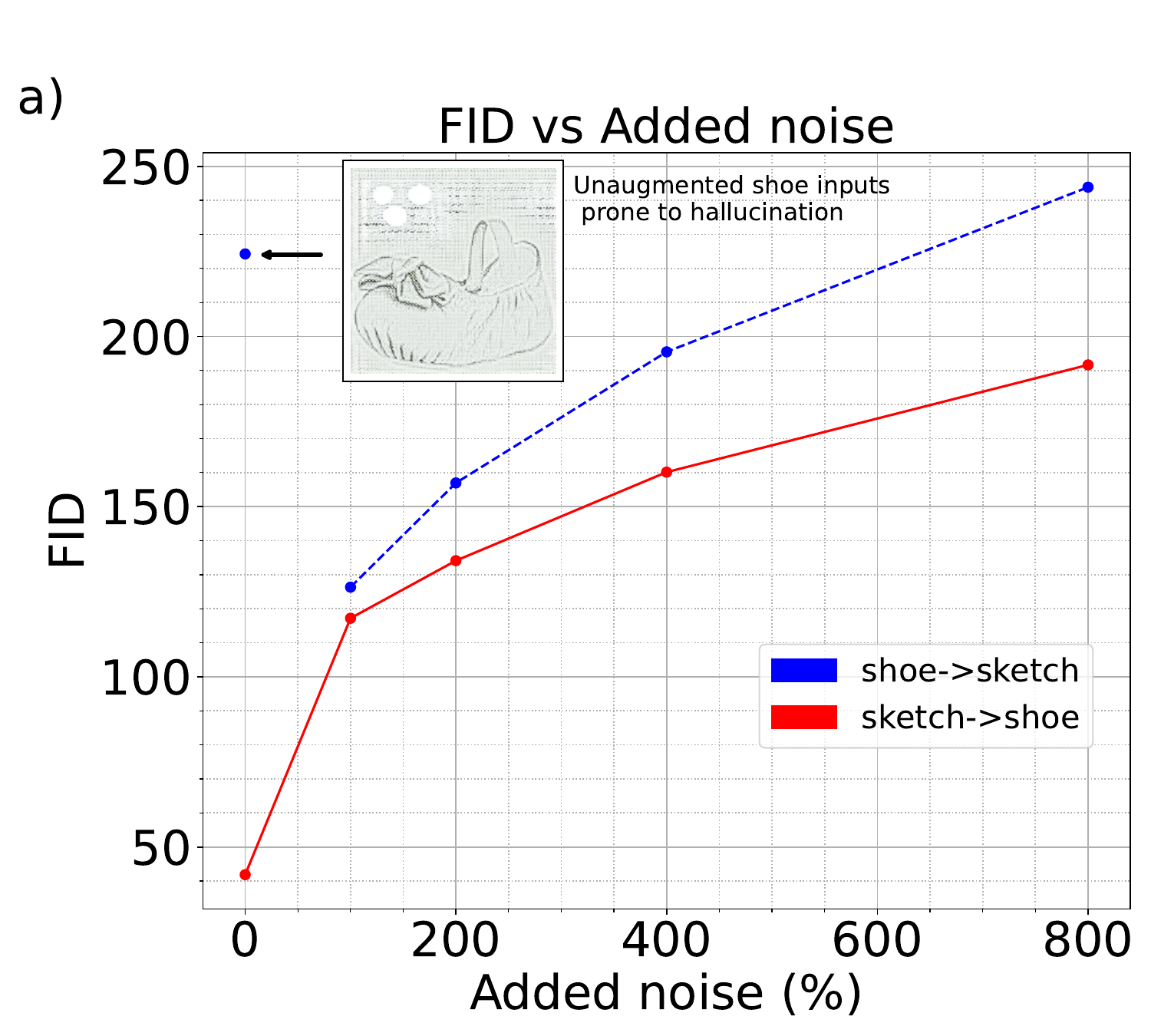}
    \includegraphics[width=.9\columnwidth]{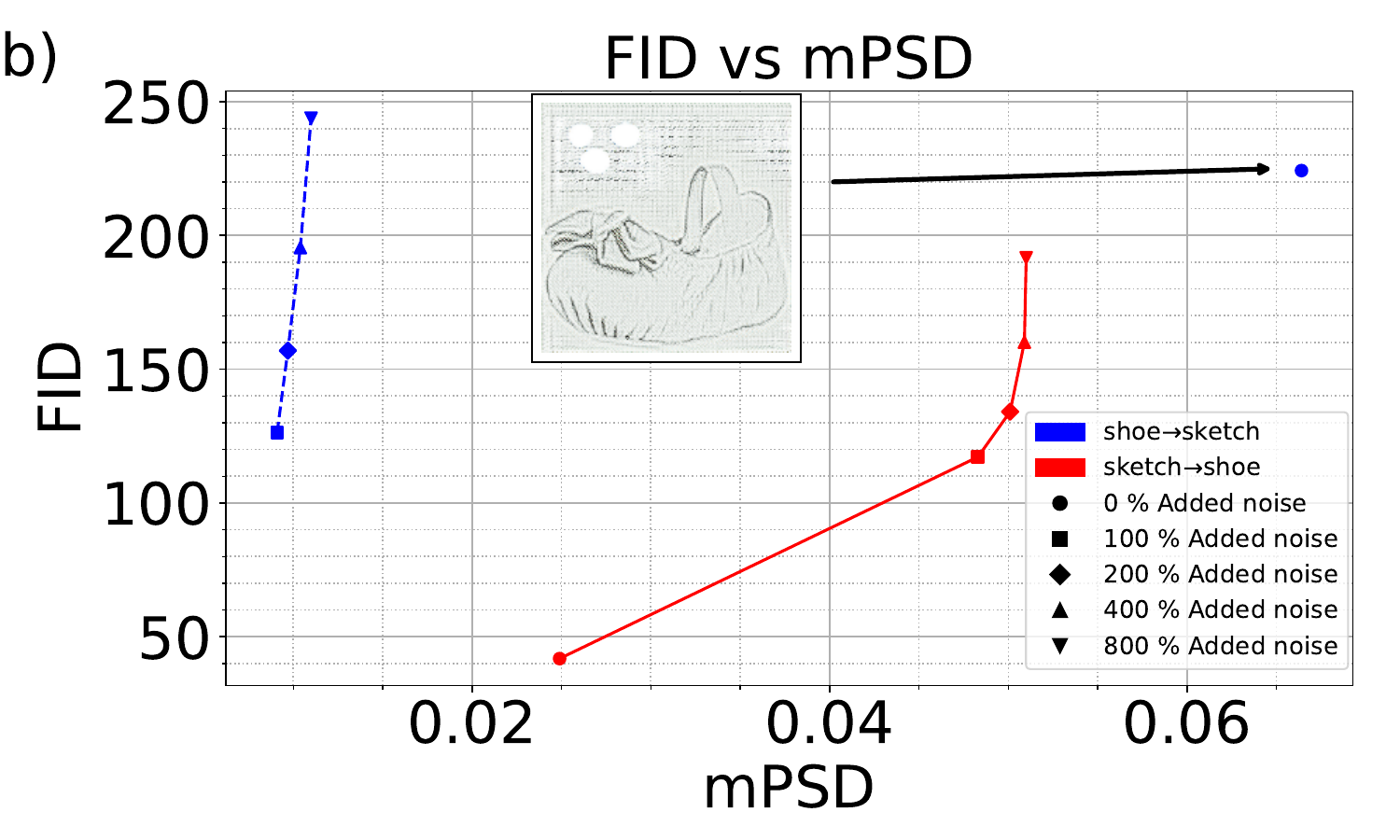}
    
    \caption{a) Plot of FID vs added noise for shoes $\leftrightarrow$ sketches (cycleGAN), showing that the FID increases with added noise for augmented images. This coarsely validates its use as an accuracy metric. Unaugmented shoe images produce poor quality outputs likely due to the model's handling of empty (low variability in pixel amplitude) regions of the input. The annotation provides an example output sketch produced from an unaugmented shoe input. Gaussian noise is added to each input image, with a mean of zero and a variance defined by the \% of the amplitude of the max pixel value in the image (horizontal axis). b) Plot of FID vs mPSD for the same tasks. We see a relationship between FID and mPSD for augmented images.}
    \label{fig:fidvsuncert_shoes}
\end{figure}

\begin{figure}
    \centering
    \includegraphics[trim={0 2.3cm 0 1.7cm},clip,width=.7\columnwidth]{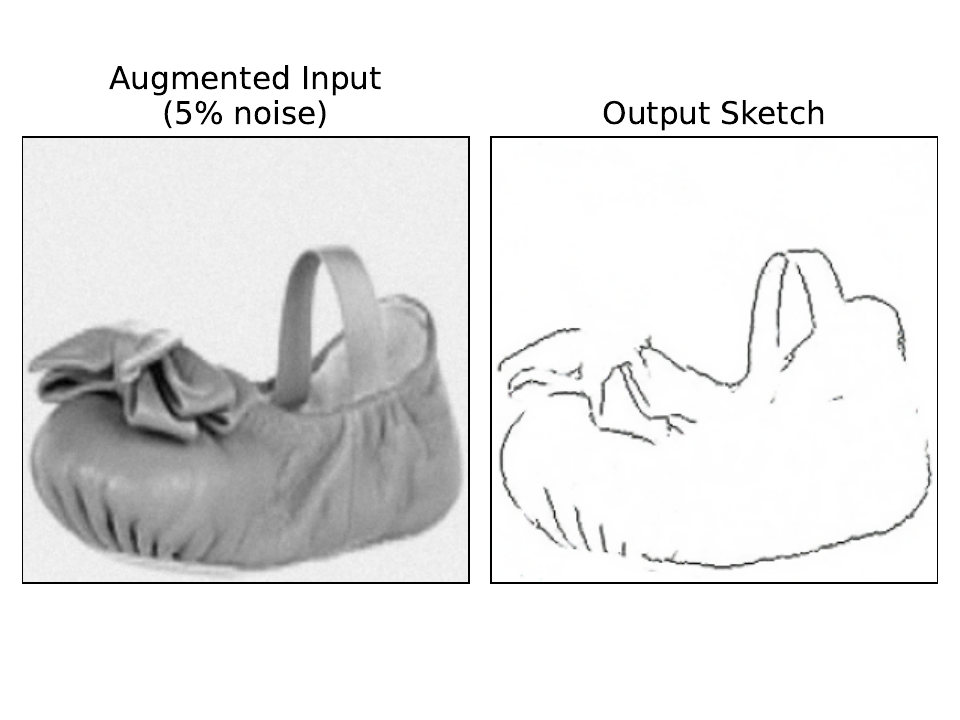}
    
    \caption{An augmented input shoe image, with its style-transferred sketch counterpart. The output does not feature the hallucinations seen in the output (annotation of Fig\ref{fig:fidvsuncert_shoes}) produced using the unaugmented version of the shoe image, despite using the same model. Adding noise removes empty (low variability in pixel amplitude) regions of the input prone to hallucinations.}
    \label{fig:shoes}
\end{figure}

For sketches $\rightarrow$ shoes, we find that the FID increases with the level of added noise, broadly validating its use as an accuracy metric, see Fig~\ref{fig:fidvsuncert_shoes}. However, for the shoes $\rightarrow$ sketches task we found that adding small amounts of noise dramatically improves performance. We hypothesise that hallucinations are prone to occur in empty (low variability in pixel amplitude) regions of the shoe images, and that adding noise can reduce instances of these artefacts. An example of adding just 5 \% noise is given in Fig~\ref{fig:shoes} (compare to the annotation in Fig~\ref{fig:fidvsuncert_shoes}). The mPSD for the shoe $\rightarrow$ sketch outputs with no noise is high. Given hallucinations are one indicator of poor performance and that, ideally, the magnitude of a well-calibrated uncertainty metric should should correlate with model performance, these results indicate that the PSD may be a useful uncertainty metric. With the exception of the no noise case, we observe the same trends of FID increasing with noise for the shoe $\rightarrow$ sketch, and the mPSD correlating with FID, indicating that the PSD may be interpreted as an uncertainty.

\subsection{cycleGAN and SynDiff evaluation: mammography tasks}
\begin{figure}[h!]
    \centering
    \includegraphics[width=\columnwidth]{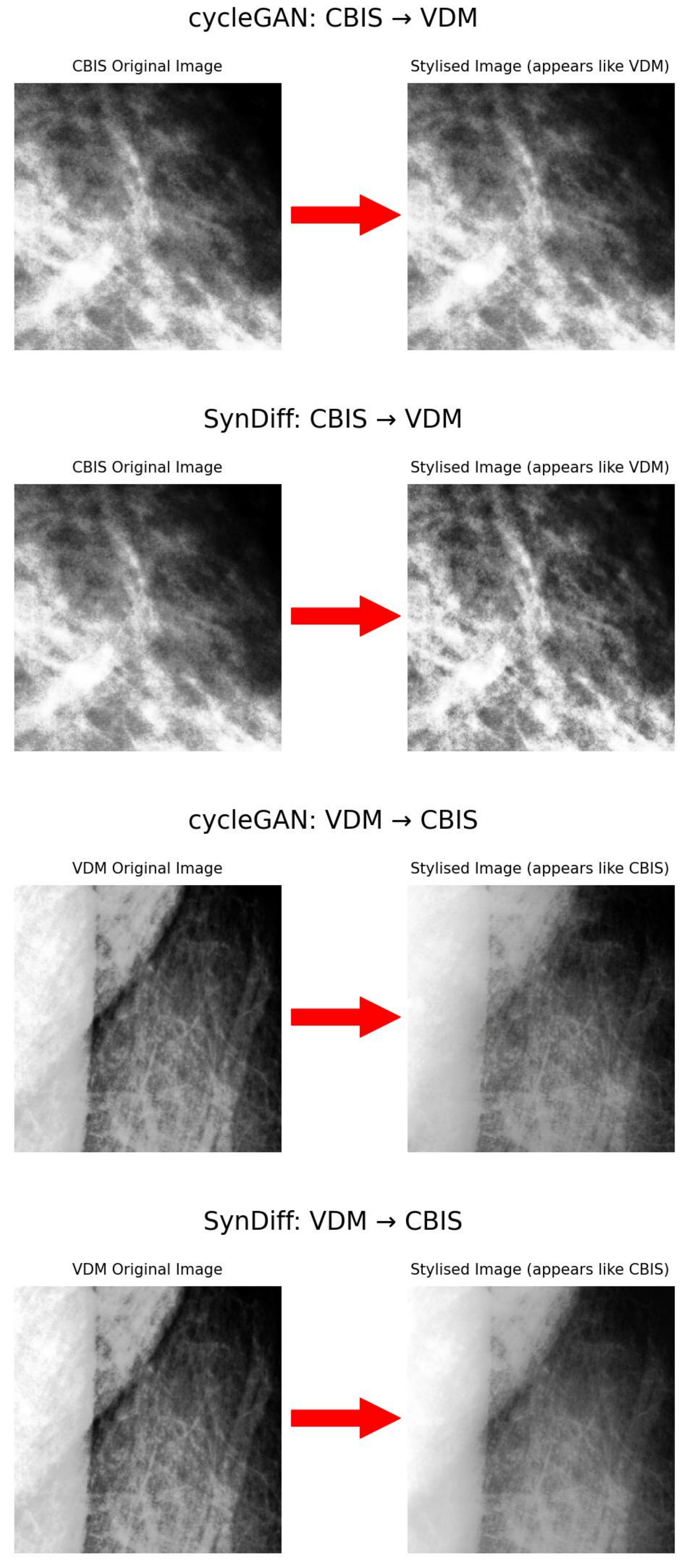}
    \caption{An example from each model for the VDM $\leftrightarrow$ CBIS task.}
    \label{fig:outputs}
\end{figure}

Some example images from the style transfer task for each model are given in Fig. \ref{fig:outputs}. Table. \ref{tab:FIDscores} shows that both the cycleGAN and SynDiff models produced style-transferred image patches that result in lower FID scores than the baseline for all tasks. We also see high CWSSIM scores, indicating that much of the structural content of the tissue has been preserved. This broadly suggests that the 
adapted source images are in greater alignment with the target domain
than the unadapted source images, and that the adapted images still depict the tissues structural features in the unadapted images.

\begin{table}[ht!]
\centering
\caption{FID (CWSSIM) scores: Mammography image tasks}
\begin{tabular}{|c|c|c|c|}
\hline
Task & Baseline FID & cycleGAN & SynDiff
\\
\hline
CMMD $\rightarrow$ CBIS & 35.2 & 32.7 ({\bf0.94}) & {\bf 15.2} (0.93)  \\
CBIS $\rightarrow$ CMMD  & 35.2 & 21.4 ({\bf0.97}) & {\bf 20.0} (0.92)\\
\hline
CMMD $\rightarrow$ VDM & 37.2 & {\bf 30.5} ({\bf0.97}) & 34.2 (0.96)\\
VDM $\rightarrow$ CMMD  & 37.2 & {\bf 23.2} ({\bf0.96}) & 24.1 (0.95) \\
\hline
VDM $\rightarrow$ CBIS & 44.8 & {\bf 18.7} ({\bf0.94}) & 19.8  ({\bf0.94}) \\
CBIS $\rightarrow$ VDM  & 44.8 & 40.4 ({\bf0.98})& {\bf 34.9} (0.94) \\
\hline
\end{tabular}
\label{tab:FIDscores}
\end{table}

\subsection{CycleGAN: detecting hallucinations with predicted uncertainties}
\begin{figure*}
    \centering
    \includegraphics[width=0.55\columnwidth]{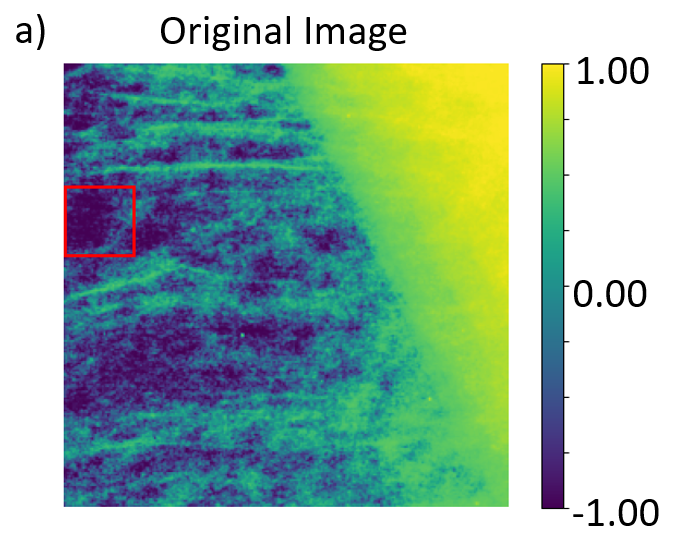}
    \includegraphics[width=0.55\columnwidth]{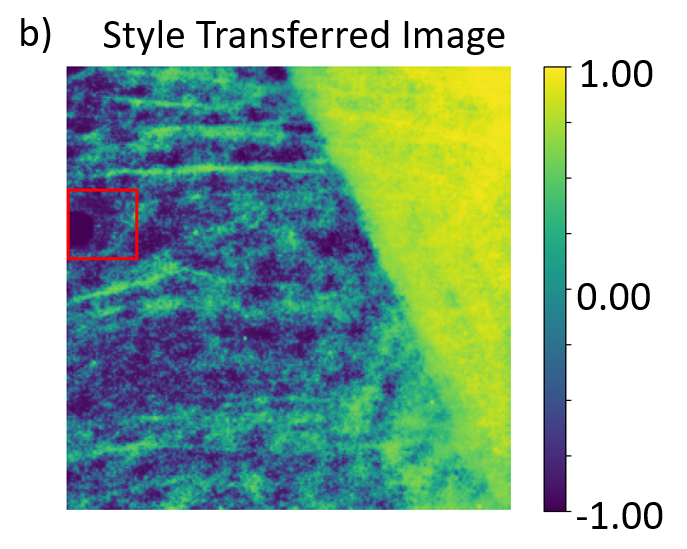}
    \includegraphics[width=0.56\columnwidth]{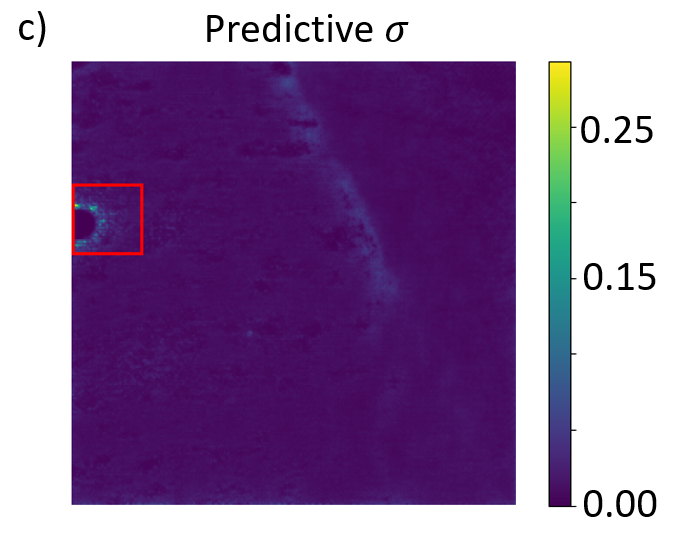}\\
    \caption{a) An unadapted CBIS image patch, where the red box indicates a sparser region of tissue. b) The style-transferred (cycleGAN) counterpart to a), that contains hallucinated features in this sparse region. c) the pixel-wise predictive standard deviation ($\sigma$). The hallucinated region has correspondingly high $\sigma$. All images are shown in viridus colour scale to improve visualisation.}
    \label{fig:hallucination}
\end{figure*}

Like shoes $\rightarrow$ sketches, for the CBIS $\rightarrow$ VDM task, we found that hallucinations were likely to occur in sparser regions of the CBIS image patches and that these regions correspond with higher PSD. We provide an example in Fig. \ref{fig:hallucination}, where the model hallucinates a well defined semi circle feature that is not present in the original image. This provides further evidence that the PSD may have the qualities we desire from an uncertainty metric. 

\subsection{CycleGAN: validating predicted uncertainties}
\begin{figure}
    \centering
    
    {\fontfamily{cmss}\selectfont \large CycleGAN: VDM $\leftrightarrow$ CBIS }\vspace{.1cm} \\
    \includegraphics[width=0.41\columnwidth]{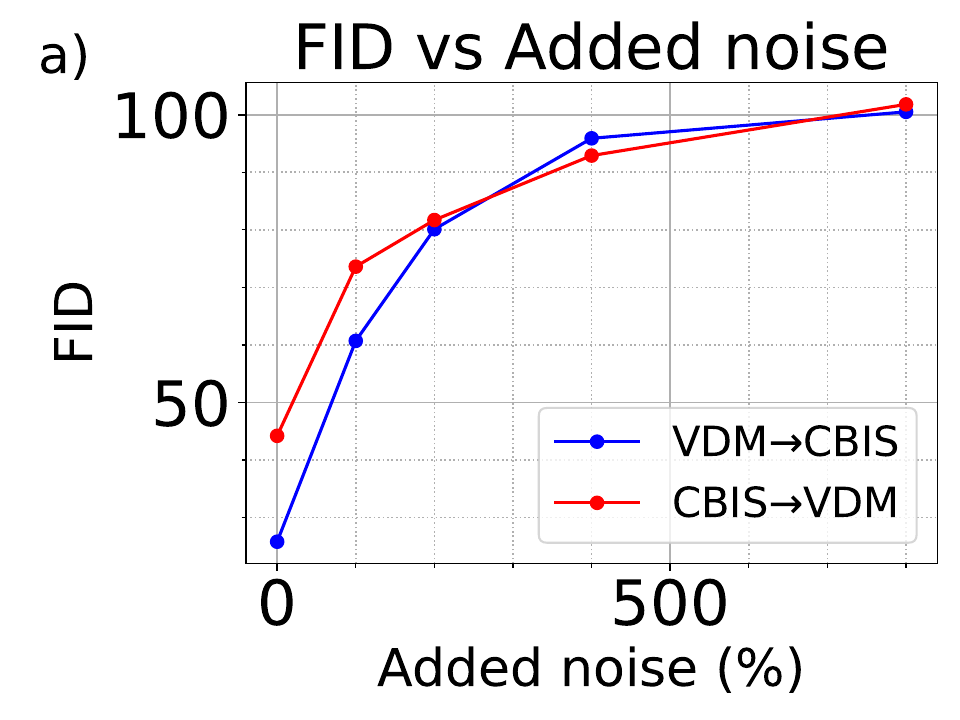}
    \includegraphics[width=0.57\columnwidth]{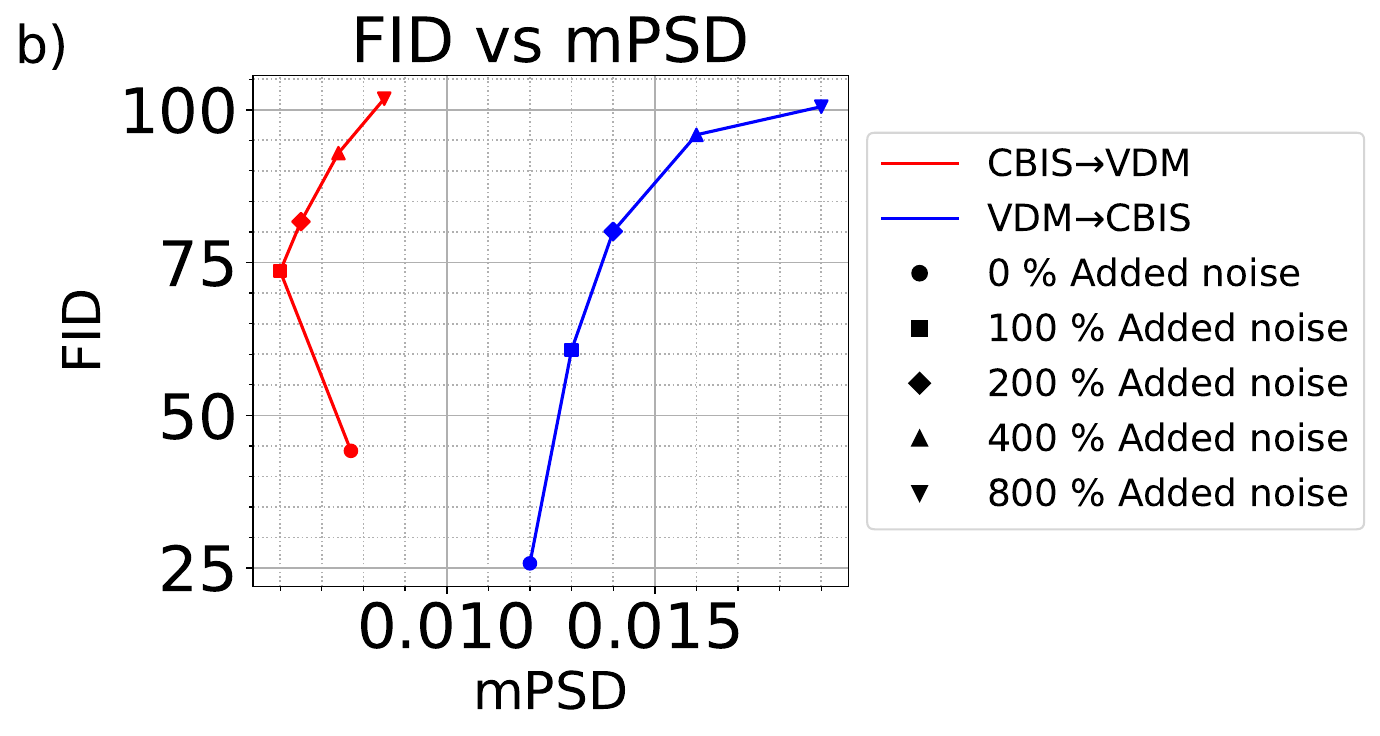}
    
    \caption{a) Plot of FID (computed from single evaluation) vs added noise for VDM $\leftrightarrow$ CBIS (cycleGAN), showing that the FID correlates with added noise. This coarsely validates its use as an accuracy metric. Gaussian noise is added to each input image, with a mean of zero and a variance defined by the \% of the amplitude of the max pixel value in the image (horizontal axis). b) Plot of the same FID vs mPSD (Monte Carlo Dropout) for VDM $\leftrightarrow$ CBIS. }
    \label{fig:fidvsuncert}
\end{figure}

While there are only a few data points, the FID does appear to positively correlate with the amount of noise added to augmented images. For CBIS $\rightarrow$ VDM, the model's performance on the unaugmented test set produced higher mPSD than for some augmented images, see Fig.~\ref{fig:fidvsuncert}. Again, we hypothesise that this is because the addition of noise prevents the occurrence of sparse regions that give rise to hallucinations. Unlike the sketches $\rightarrow$ shoes task, the FID for the outputs produced from augmented inputs is higher than for those produced from the unaugmented inputs. We hypothesis this is because the hallucinations in an image tend to be less frequent and more localised compared to those in the shoes $\rightarrow$ sketches outputs, possibly due to our patch inclusion criteria of 99 \% tissue. 

The FID appears to plateau with increasing amounts of noise for VDM $\rightarrow$ CBIS, (see Fig.~\ref{fig:fidvsuncert}). This suggests that while adding noise does affect the network's ability to extract features from the images, once finer details are uninterpretable, the network's interpretation of image contents does not appear change significantly with large amounts of added noise. 

We also find that the FID correlates with the mPSD, indicating that it does encode some information about accuracy, which is an ideal property of an estimated uncertainty. Though, given the FID only allows us to assess model accuracy with low precision, we can not provide further comment on how well the PSD represents uncertainty.

\subsection{SynDiff: validating predicted uncertainties}
\begin{figure}
    \centering
    
    {\fontfamily{cmss}\selectfont \large SynDiff: VDM $\leftrightarrow$ CBIS }\vspace{.1cm} \\
    \includegraphics[width=0.41\columnwidth]{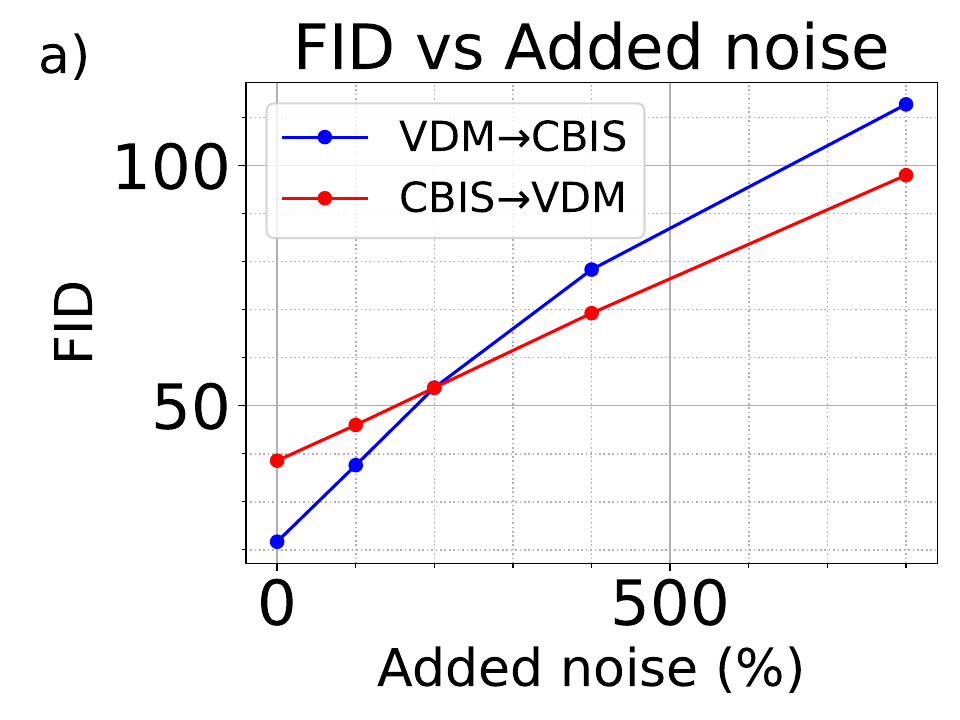}
    \includegraphics[width=0.57\columnwidth]{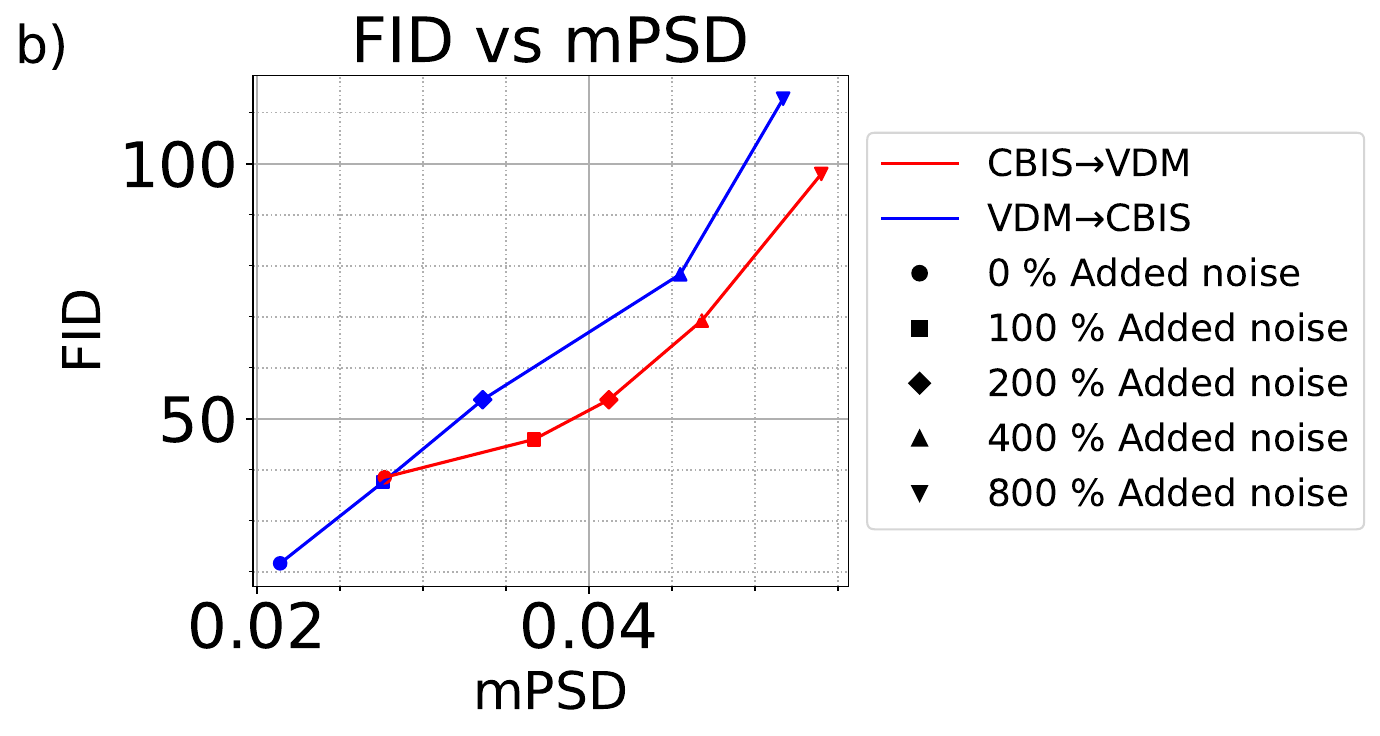}
    
    \caption{a) Plot of the mean FID computed over the outputs of five models vs added noise for VDM $\leftrightarrow$ CBIS (SynDiff). b) Plot of the FID vs mPSD (Deep Ensembles) for VDM $\leftrightarrow$ CBIS. }
    \label{fig:syndiff_fidvsuncert}
\end{figure}

In Fig.~\ref{fig:syndiff_fidvsuncert} we find that the FID appears to correlate with the amount of added noise to images, and that the mPSD correlates with their corresponding FID values. These results indicate that our evaluation protocol can be used to assess calibration for other kinds of generative models, and uncertainty quantification techniques.

\section{Discussion}

Both the cycleGAN and SynDiff models produced high quality outputs for all mammography-image tasks. We applied uncertainty quantification for both models, and demonstrated that at least for one task (CBIS $\rightarrow$ VDM) how images of the predictive standard deviation can be used to highlight regions likely to contain hallucinations. This demonstrates one practical benefit of incorporating uncertainty quantification into generative models. 

We also proposed and implemented a scheme for assessing uncertainty calibration. We have shown examples for both GAN and diffusion based models, and for two uncertainty quantification techniques: Monte Carlo Dropout and deep ensembles. When comparing across the results produced from images augmented with noise, increasing amounts of noise increased the FID for both models, validating its use as an accuracy metric (this is akin to similar results reported in \cite{buzuti2023frechet}). We also found that for augmented images, these FID scores suggest a correlation with changes in mPSD, suggesting that the PSD exhibits properties of a well-formulated uncertainty metric  (i.e. it correlates with accuracy). Therefore, we suggest that this procedure can be used to determine whether the estimated PSD may be interpreted as an uncertainty in the absence of ground truths.

While there are only a few data points, we found that FID and mPSD produced from the ensembling of SynDiff models (Fig. \ref{fig:syndiff_fidvsuncert}) did not appear to plateau with added noise when compared to the cycleGAN experiments (Fig. \ref{fig:fidvsuncert}), and produced larger mPSD values. Future work will consider how much of this behaviour is due to the use of a diffusion model, or deep ensembles.

With that said, our evaluation protocol is limited in several ways. For example, we only condition our assessment of uncertainty calibration on the amount of noise added to the images. This contrasts with the typical approach of conditioning on the magnitude of the estimated uncertainties \cite{pernot2023calibration}. While this provides a data-efficient approach to conditional calibration, the ramifications are as follows. When assessing calibration quality, we hope to acquire an idea of how well calibrated the uncertainties are when produced from a range of different test examples. This helps us determine how well the model's capacity to predict uncertainties generalises to a diverse set of unseen test examples. Conditional calibration techniques aim to assess this by assessing calibration for individual bins, each containing a distinct population of examples (e.g. binned by magnitude of predicted uncertainties). Yet here, we only assess this over augmented versions of a fixed set of test examples, and with few datapoints. So the calibration quality one might derive using this scheme may change considerably with the inclusion of more data points, or not be representative of that computed for a diverse set of unseen test examples. With that said, given constraints on the number of available images in medical imaging, it is challenging to develop local calibration techniques without large amounts of data or ground truths. Future work could consider the inclusion of more datapoints for a more thorough validation of any potential correlations.

There are other way we could extend our study. For the cycleGAN, we didn't experiment with various dropout rates, or MC samples due to constraints on memory. Additionally, we were limited to 5 models for the ensembling of the SynDiff models due to computational expense. Given these hyperparameters are known to affect the quality of estimated uncertainties, this limits the extent to which we could assess calibration quality.

Furthermore, the evaluation scheme relies on the assumption that the models do not generalise their performance well to progressively noisier versions of the test set. This may not be the case for some models/tasks, and more appropriate augmentation schemes may have to be applied in other applications. For example, we also applied our evaluation scheme on a translation task involving non-medical image data (shoes $\leftrightarrow$ sketches) with the cycleGAN, with the aim of providing some indication of the generalisability of our approach. However, we found that the cycleGAN was prone to hallucinations when processing examples with empty space, and that adding small amounts of noise could actually improve performance relative to the unaugmented case. While adding subsequently larger amounts of noise then degraded performance allowing us to use our evaluation protocol, it nonetheless highlights the importance of investigating model behaviour with increasing augmentation strength. Indeed, this experiment with non-natural images showed that our approach can be used for tasks outside of medical imaging. 

\section{Conclusion}
We have proposed and implemented a scheme for validating uncertainties estimated from unpaired image-to-image translation models. While we have shown promising results on tasks considering both medical and non-medical images, and for both GAN and diffusion-based frameworks (each using different uncertainty quantification techniques), it only assesses calibration quality over augmented versions of a fixed set of test examples. This ultimately limits whether the calibration derived from the test set will generalise to diverse sets of unseen test examples. Nonetheless, our scheme can provide some indication as to whether estimates of the PSD may be interpreted as uncertainties. Ultimately, this is a first step towards developing more thorough test/evaluation schemes.

\section*{Acknowledgment}
The project 22HLT05 MAIBAI has received funding from the European Partnership on Metrology, co-financed from the European Union’s Horizon Europe Research and Innovation Programme and by the Participating States. Funding for the UK partners was provided by Innovate UK under the Horizon Europe Guarantee Extension.

\bibliographystyle{unsrt}  
\bibliography{references}

\begin{thebibliography}{10}

\bibitem{arnold2022current}
Melina Arnold, Eileen Morgan, Harriet Rumgay, Allini Mafra, Deependra Singh, Mathieu Laversanne, Jerome Vignat, Julie~R Gralow, Fatima Cardoso, Sabine Siesling, et~al.
\newblock Current and future burden of breast cancer: Global statistics for 2020 and 2040.
\newblock {\em The Breast}, 66:15--23, 2022.

\bibitem{carlson2003treatment}
Robert~W Carlson, Benjamin~O Anderson, Rakesh Chopra, Alexandru~E Eniu, Raimund Jakesz, Richard~R Love, Riccardo Masetti, and Gilberto Schwartsmann.
\newblock Treatment of breast cancer in countries with limited resources.
\newblock {\em The breast journal}, 9:S67--S74, 2003.

\bibitem{mckinney2020international}
Scott~Mayer McKinney, Marcin Sieniek, Varun Godbole, Jonathan Godwin, Natasha Antropova, Hutan Ashrafian, Trevor Back, Mary Chesus, Greg~S Corrado, Ara Darzi, et~al.
\newblock International evaluation of an ai system for breast cancer screening.
\newblock {\em Nature}, 577(7788):89--94, 2020.

\bibitem{paul2016deep}
Rahul Paul, Samuel~H Hawkins, Yoganand Balagurunathan, Matthew Schabath, Robert~J Gillies, Lawrence~O Hall, and Dmitry~B Goldgof.
\newblock Deep feature transfer learning in combination with traditional features predicts survival among patients with lung adenocarcinoma.
\newblock {\em Tomography}, 2(4):388--395, 2016.

\bibitem{bai2021don}
Yu~Bai, Song Mei, Huan Wang, and Caiming Xiong.
\newblock Don’t just blame over-parametrization for over-confidence: Theoretical analysis of calibration in binary classification.
\newblock In {\em International conference on machine learning}, pages 566--576. PMLR, 2021.

\bibitem{saranya2023systematic}
A~Saranya and R~Subhashini.
\newblock A systematic review of explainable artificial intelligence models and applications: Recent developments and future trends.
\newblock {\em Decision analytics journal}, 7:100230, 2023.

\bibitem{pertuz2023saliency}
Said Pertuz, David Ortega, {\'E}rika Suarez, William Cancino, Gerson Africano, Irina Rinta-Kiikka, Otso Arponen, Sara Paris, and Alfonso Lozano.
\newblock Saliency of breast lesions in breast cancer detection using artificial intelligence.
\newblock {\em Scientific Reports}, 13(1):20545, 2023.

\bibitem{oza2022image}
Parita Oza, Paawan Sharma, Samir Patel, Festus Adedoyin, and Alessandro Bruno.
\newblock Image augmentation techniques for mammogram analysis.
\newblock {\em journal of imaging}, 8(5):141, 2022.

\bibitem{kouw2019review}
Wouter~M Kouw and Marco Loog.
\newblock A review of domain adaptation without target labels.
\newblock {\em IEEE transactions on pattern analysis and machine intelligence}, 43(3):766--785, 2019.

\bibitem{costa2019data}
Arthur~C Costa, Helder~CR Oliveira, and Marcelo~AC Vieira.
\newblock Data augmentation: Effect in deep convolutional neural network for the detection of architectural distortion in digital mammography.
\newblock 2019.

\bibitem{ntelemis_generic_2024}
Foivos Ntelemis, Yaochu Jin, and Spencer~A. Thomas.
\newblock A {Generic} {Self}-{Supervised} {Framework} of {Learning} {Invariant} {Discriminative} {Features}.
\newblock {\em IEEE Transactions on Neural Networks and Learning Systems}, 35(9):12938--12952, September 2024.
\newblock Conference Name: IEEE Transactions on Neural Networks and Learning Systems.

\bibitem{hoyez2022unsupervised}
Henri Hoyez, C{\'e}dric Schockaert, Jason Rambach, Bruno Mirbach, and Didier Stricker.
\newblock Unsupervised image-to-image translation: A review.
\newblock {\em Sensors}, 22(21):8540, 2022.

\bibitem{mariapun2015ethnic}
Shivaani Mariapun, Jingmei Li, Cheng~Har Yip, Nur Aishah~Mohd Taib, and Soo-Hwang Teo.
\newblock Ethnic differences in mammographic densities: an asian cross-sectional study.
\newblock {\em PloS one}, 10(2):e0117568, 2015.

\bibitem{wu2018conditional}
Eric Wu, Kevin Wu, David Cox, and William Lotter.
\newblock Conditional infilling gans for data augmentation in mammogram classification.
\newblock In {\em Image Analysis for Moving Organ, Breast, and Thoracic Images: Third International Workshop, RAMBO 2018, Fourth International Workshop, BIA 2018, and First International Workshop, TIA 2018, Held in Conjunction with MICCAI 2018, Granada, Spain, September 16 and 20, 2018, Proceedings 3}, pages 98--106. Springer, 2018.

\bibitem{wang2020mr}
Sheng Wang, Jiayu Huo, Xi~Ouyang, Jifei Che, Zhong Xue, Dinggang Shen, Qian Wang, and Jie-Zhi Cheng.
\newblock mr nst: Multi-resolution and multi-reference neural style transfer for mammography.
\newblock In {\em International Workshop on PRedictive Intelligence In MEdicine}, pages 169--177. Springer, 2020.

\bibitem{mcnaughton2023machine}
Jake McNaughton, Justin Fernandez, Samantha Holdsworth, Benjamin Chong, Vickie Shim, and Alan Wang.
\newblock Machine learning for medical image translation: A systematic review.
\newblock {\em Bioengineering}, 10(9):1078, 2023.

\bibitem{jung2018inferring}
Merel~M Jung, Bram van~den Berg, Eric Postma, and Willem Huijbers.
\newblock Inferring pet from mri with pix2pix.
\newblock In {\em Benelux Conference on Artificial Intelligence}, volume~9, 2018.

\bibitem{fernandez2021improving}
Alvaro Fernandez-Quilez, Steinar~Valle Larsen, Morten Goodwin, Thor~Ole Gulsrud, Svein~Reidar Kjosavik, and Ketil Oppedal.
\newblock Improving prostate whole gland segmentation in t2-weighted mri with synthetically generated data.
\newblock In {\em 2021 IEEE 18th International Symposium on Biomedical Imaging (ISBI)}, pages 1915--1919. IEEE, 2021.

\bibitem{zhu2017unpaired}
Jun-Yan Zhu, Taesung Park, Phillip Isola, and Alexei~A Efros.
\newblock Unpaired image-to-image translation using cycle-consistent adversarial networks.
\newblock In {\em Proceedings of the IEEE international conference on computer vision}, pages 2223--2232, 2017.

\bibitem{ozbey2023unsupervised}
Muzaffer {\"O}zbey, Onat Dalmaz, Salman~UH Dar, Hasan~A Bedel, {\c{S}}aban {\"O}zturk, Alper G{\"u}ng{\"o}r, and Tolga {\c{C}}ukur.
\newblock Unsupervised medical image translation with adversarial diffusion models.
\newblock {\em IEEE Transactions on Medical Imaging}, 2023.

\bibitem{yang2023diffusion}
Ling Yang, Zhilong Zhang, Yang Song, Shenda Hong, Runsheng Xu, Yue Zhao, Wentao Zhang, Bin Cui, and Ming-Hsuan Yang.
\newblock Diffusion models: A comprehensive survey of methods and applications.
\newblock {\em ACM Computing Surveys}, 56(4):1--39, 2023.

\bibitem{cha2020evaluation}
Kenny~H Cha, Nicholas Petrick, Aria Pezeshk, Christian~G Graff, Diksha Sharma, Andreu Badal, and Berkman Sahiner.
\newblock Evaluation of data augmentation via synthetic images for improved breast mass detection on mammograms using deep learning.
\newblock {\em Journal of Medical Imaging}, 7(1):012703--012703, 2020.

\bibitem{pernot2023calibration}
Pascal Pernot.
\newblock Calibration in machine learning uncertainty quantification: beyond consistency to target adaptivity.
\newblock {\em APL Machine Learning}, 1(4), 2023.

\bibitem{kendall2017uncertainties}
Alex Kendall and Yarin Gal.
\newblock What uncertainties do we need in bayesian deep learning for computer vision?
\newblock {\em Advances in neural information processing systems}, 30, 2017.

\bibitem{upadhyay2021uncertainty}
Uddeshya Upadhyay, Viswanath~P Sudarshan, and Suyash~P Awate.
\newblock Uncertainty-aware gan with adaptive loss for robust mri image enhancement.
\newblock In {\em Proceedings of the IEEE/CVF International Conference on Computer Vision}, pages 3255--3264, 2021.

\bibitem{upadhyay2021robustness}
Uddeshya Upadhyay, Yanbei Chen, and Zeynep Akata.
\newblock Robustness via uncertainty-aware cycle consistency.
\newblock {\em Advances in neural information processing systems}, 34:28261--28273, 2021.

\bibitem{karthik2023uncertainty}
Enamundram~Naga Karthik, Farida Cheriet, and Catherine Laporte.
\newblock Uncertainty estimation in unsupervised mr-ct synthesis of scoliotic spines.
\newblock {\em IEEE Open Journal of Engineering in Medicine and Biology}, 2023.

\bibitem{palakkadavath2021bayesian}
Ragja Palakkadavath and PK~Srijith.
\newblock Bayesian generative adversarial nets with dropout inference.
\newblock In {\em Proceedings of the 3rd ACM India Joint International Conference on Data Science \& Management of Data (8th ACM IKDD CODS \& 26th COMAD)}, pages 92--100, 2021.

\bibitem{tiao2018cycle}
Louis~C Tiao, Edwin~V Bonilla, and Fabio Ramos.
\newblock Cycle-consistent adversarial learning as approximate bayesian inference.
\newblock {\em arXiv preprint arXiv:1806.01771}, 2018.

\bibitem{you2020bayesian}
Haoran You, Yu~Cheng, Tianheng Cheng, Chunliang Li, and Pan Zhou.
\newblock Bayesian cycle-consistent generative adversarial networks via marginalizing latent sampling.
\newblock {\em IEEE Transactions on Neural Networks and Learning Systems}, 32(10):4389--4403, 2020.

\bibitem{RanWang2022}
Haoran You, Yu~Cheng, Tianheng Cheng, Chunliang Li, and Pan Zhou.
\newblock Bayesian cycle-consistent generative adversarial networks via marginalizing latent sampling.
\newblock {\em IEEE Transactions on Neural Networks and Learning Systems}, 32(10):4389--4403, 2020.

\bibitem{galapon2024feasibility}
Arthur~Villanueva Galapon~Jr, Adrian Thummerer, Johannes~Albertus Langendijk, Dirk Wagenaar, and Stefan Both.
\newblock Feasibility of monte carlo dropout-based uncertainty maps to evaluate deep learning-based synthetic cts for adaptive proton therapy.
\newblock {\em Medical Physics}, 51(4):2499--2509, 2024.

\bibitem{kou2023bayesdiff}
Siqi Kou, Lei Gan, Dequan Wang, Chongxuan Li, and Zhijie Deng.
\newblock Bayesdiff: Estimating pixel-wise uncertainty in diffusion via bayesian inference.
\newblock {\em arXiv preprint arXiv:2310.11142}, 2023.

\bibitem{xu2024bayesian}
Haiyang Xu, Yu~Lei, Zeyuan Chen, Xiang Zhang, Yue Zhao, Yilin Wang, and Zhuowen Tu.
\newblock Bayesian diffusion models for 3d shape reconstruction.
\newblock In {\em Proceedings of the IEEE/CVF Conference on Computer Vision and Pattern Recognition}, pages 10628--10638, 2024.

\bibitem{neumeier2024reliable}
Marion Neumeier, Sebastian Dorn, Michael Botsch, and Wolfgang Utschick.
\newblock Reliable trajectory prediction and uncertainty quantification with conditioned diffusion models.
\newblock In {\em Proceedings of the IEEE/CVF Conference on Computer Vision and Pattern Recognition}, pages 3461--3470, 2024.

\bibitem{lakshminarayanan2017simple}
Balaji Lakshminarayanan, Alexander Pritzel, and Charles Blundell.
\newblock Simple and scalable predictive uncertainty estimation using deep ensembles.
\newblock {\em Advances in neural information processing systems}, 30, 2017.

\bibitem{shu2024zero}
Dule Shu and Amir~Barati Farimani.
\newblock Zero-shot uncertainty quantification using diffusion probabilistic models.
\newblock {\em arXiv preprint arXiv:2408.04718}, 2024.

\bibitem{berryshedding}
Lucas Berry, Axel Brando, and David Meger.
\newblock Shedding light on large generative networks: Estimating epistemic uncertainty in diffusion models.
\newblock In {\em The 40th Conference on Uncertainty in Artificial Intelligence}.

\bibitem{ekmekci2023quantifying}
Canberk Ekmekci and Mujdat Cetin.
\newblock Quantifying generative model uncertainty in posterior sampling methods for computational imaging.
\newblock In {\em NeurIPS 2023 Workshop on Deep Learning and Inverse Problems}, 2023.

\bibitem{chan2024hyper}
Matthew~A Chan, Maria~J Molina, and Christopher~A Metzler.
\newblock Hyper-diffusion: Estimating epistemic and aleatoric uncertainty with a single model.
\newblock {\em arXiv preprint arXiv:2402.03478}, 2024.

\bibitem{heusel2017gans}
Martin Heusel, Hubert Ramsauer, Thomas Unterthiner, Bernhard Nessler, and Sepp Hochreiter.
\newblock Gans trained by a two time-scale update rule converge to a local nash equilibrium.
\newblock {\em Advances in neural information processing systems}, 30, 2017.

\bibitem{deshpande2024report}
Rucha Deshpande, Varun~A Kelkar, Dimitrios Gotsis, Prabhat Kc, Rongping Zeng, Kyle~J Myers, Frank~J Brooks, and Mark~A Anastasio.
\newblock Report on the aapm grand challenge on deep generative modeling for learning medical image statistics.
\newblock {\em ArXiv}, 2024.

\bibitem{nguyen2023vindr}
Hieu~T Nguyen, Ha~Q Nguyen, Hieu~H Pham, Khanh Lam, Linh~T Le, Minh Dao, and Van Vu.
\newblock Vindr-mammo: A large-scale benchmark dataset for computer-aided diagnosis in full-field digital mammography.
\newblock {\em Scientific Data}, 10(1):277, 2023.

\bibitem{cai2023online}
Hongmin Cai, Jinhua Wang, Tingting Dan, Jiao Li, Zhihao Fan, Weiting Yi, Chunyan Cui, Xinhua Jiang, and Li~Li.
\newblock An online mammography database with biopsy confirmed types.
\newblock {\em Scientific Data}, 10(1):123, 2023.

\bibitem{sawyer2016curated}
Rebecca Sawyer-Lee, Francisco Gimenez, Assaf Hoogi, and Daniel Rubin.
\newblock Curated breast imaging subset of digital database for screening mammography (cbis-ddsm).
\newblock {\em (No Title)}, 2016.

\bibitem{ostu1979threshold}
Nobuyuki Ostu.
\newblock A threshold selection method from gray-level histograms.
\newblock {\em IEEE Trans SMC}, 9:62, 1979.

\bibitem{saad2024early}
Muhammad~Muneeb Saad, Mubashir~Husain Rehmani, and Ruairi O'Reilly.
\newblock Early stopping criteria for training generative adversarial networks in biomedical imaging.
\newblock {\em arXiv preprint arXiv:2405.20987}, 2024.

\bibitem{cwssim}
Mehul~P. Sampat, Zhou Wang, Shalini Gupta, Alan~Conrad Bovik, and Mia~K. Markey.
\newblock Complex wavelet structural similarity: A new image similarity index.
\newblock {\em IEEE Transactions on Image Processing}, 18(11):2385 -- 2401, 2009.

\bibitem{graf2023denoising}
Robert Graf, Joachim Schmitt, Sarah Schlaeger, Hendrik~Kristian M{\"o}ller, Vasiliki Sideri-Lampretsa, Anjany Sekuboyina, Sandro~Manuel Krieg, Benedikt Wiestler, Bjoern Menze, Daniel Rueckert, et~al.
\newblock Denoising diffusion-based mri to ct image translation enables automated spinal segmentation.
\newblock {\em European Radiology Experimental}, 7(1):70, 2023.

\bibitem{liang2022image}
Zhaohui Liang, Jimmy~Xiangji Huang, and Sameer Antani.
\newblock Image translation by ad cyclegan for covid-19 x-ray images: a new approach for controllable gan.
\newblock {\em Sensors}, 22(24):9628, 2022.

\bibitem{abdusalomov2023evaluating}
Akmalbek~Bobomirzaevich Abdusalomov, Rashid Nasimov, Nigorakhon Nasimova, Bahodir Muminov, and Taeg~Keun Whangbo.
\newblock Evaluating synthetic medical images using artificial intelligence with the gan algorithm.
\newblock {\em Sensors}, 23(7):3440, 2023.

\bibitem{7404021}
Maxine Tan, Bin Zheng, Joseph~K. Leader, and David Gur.
\newblock Association between changes in mammographic image features and risk for near-term breast cancer development.
\newblock {\em IEEE Transactions on Medical Imaging}, 35(7):1719--1728, 2016.

\bibitem{super-resolution}
Kensuke Umehara, Junko Ota, and Takayuki Ishida.
\newblock Super-resolution imaging of mammograms based on the super-resolution convolutional neural network.
\newblock {\em Open Journal of Medical Imaging}, 7(4):180--195, 2017.

\bibitem{Gan-baseddataaugmentation}
Yuliana Jiménez-Gaona, Diana Carrión-Figueroa, Vasudevan Lakshminarayanan, and María José Rodríguez-Álvarez.
\newblock Gan-based data augmentation to improve breast ultrasound and mammography mass classification.
\newblock {\em Biomedical Signal Processing and Control}, 94, 2024.

\bibitem{1284395}
Zhou Wang, A.C. Bovik, H.R. Sheikh, and E.P. Simoncelli.
\newblock Image quality assessment: from error visibility to structural similarity.
\newblock {\em IEEE Transactions on Image Processing}, 13(4):600--612, 2004.

\bibitem{buzuti2023frechet}
Lucas~F Buzuti and Carlos~E Thomaz.
\newblock Fr{\'e}chet autoencoder distance: a new approach for evaluation of generative adversarial networks.
\newblock {\em Computer Vision and Image Understanding}, 235:103768, 2023.

\bibitem{fort2019deep}
Stanislav Fort, Huiyi Hu, and Balaji Lakshminarayanan.
\newblock Deep ensembles: A loss landscape perspective.
\newblock {\em arXiv preprint arXiv:1912.02757}, 2019.

\end{thebibliography}

\end{document}